\documentclass[a4paper,conference]{IEEEtran}
\IEEEoverridecommandlockouts
\usepackage{cite}
\usepackage{amsmath,amssymb,amsfonts}
\usepackage{algorithm}
\usepackage{algorithmic}
\usepackage{orcidlink}
\usepackage{graphicx}
\usepackage{textcomp}

\usepackage{xcolor}
\def\BibTeX{{\rm B\kern-.05em{\sc i\kern-.025em b}\kern-.08em
    T\kern-.1667em\lower.7ex\hbox{E}\kern-.125emX}}

\begin{document}

\title{Adaptive Three Layer Hybrid Reconfigurable Intelligent Surface for 6G Wireless Communication: Trade-offs and Performance}
\author{
    \IEEEauthorblockA{
        Rashed Hasan Ratul\textsuperscript{1,*\thanks{Corresponding author. Email: rashedhasan@iut-dhaka.edu}}\orcidlink{0000-0002-5675-9214},
        Muhammad Iqbal\textsuperscript{2}\orcidlink{0000-0003-1603-0899}, Tabinda Ashraf\textsuperscript{2}\orcidlink{0000-0001-5789-8067},
        Jen-Yi Pan\textsuperscript{2}\orcidlink{0000-0003-0365-5996},
        Yi-Han Wang\textsuperscript{3}\orcidlink{0009-0002-6255-3551},
        Shao-Yu Lien\textsuperscript{4}\orcidlink{0000-0002-4347-2871}
    }
    \IEEEauthorblockA{
        \textsuperscript{1}Department of Electrical and Electronic Engineering,\\ Islamic University of Technology (IUT), Dhaka, Bangladesh\\
        \textsuperscript{2}Department of Communication Engineering, \textsuperscript{3}Department of Computer Science and Information Engineering,\\
        National Chung Cheng University (CCU), Chiayi, Taiwan\\
        \textsuperscript{4}Institute of Intelligent Systems, College of Artificial Intelligence,\\ National Yang Ming Chiao Tung University (NYCU), Tainan, Taiwan
    }
}

\maketitle

\begin{abstract}
A potential candidate technology for the development of future 6G networks has been recognized as Reconfigurable Intelligent Surface (RIS). However, due to the variation in radio link quality, traditional passive RISs only accomplish a minimal signal gain in situations with strong direct links between user equipment (UE) and base station (BS). In order to get over this fundamental restriction of smaller gain, the idea of active RISs might be a suitable solution. In contrast to current passive RIS, which simply reflects and directs signals without any additional amplification, active RISs have the ability to enhance reflected signals by the incorporation of amplifiers inside its elements. However, with additional amplifiers, apart from the relatively complex attributes of RIS-assisted arrangements, the additional energy consumption of such technologies is often disregarded. So, there might be a tradeoff between the additional energy consumption for the RIS technologies and the overall gain acquired by deploying this potential advancement. The objective of this work is to provide a primary idea of a three-layer hybrid RIS-assisted configuration that is responsive to both active and passive RIS, as well as an additional dormant or inactive state. The single RIS structure should be capable of adjusting its overall configuration in response to fluctuations in transmit power and radio link quality. 
Furthermore, our fabricated passive RIS-assisted structure verifies a portion of the proposed idea, with simulations highlighting its advantages over standalone passive or active RIS-assisted technologies.

\end{abstract}

\smallskip
\smallskip

\section{Introduction}

Wireless communications were typically regarded as non-configurable from the first generation (1G) through the fifth generation (5G) \cite{zhi2022active}. Reconfigurable Intelligent Surface (RIS) have recently been proposed as a result of advancements in meta-materials for the purpose of dynamically regulating wireless communication channels to obtain enhanced communication performance \cite{b1}. An RIS, in particular, can be regarded as an array made up of a massive quantity of passive components which reflect electromagnetic waves in a way which alters the direction wireless signals propagate. A significant benefit of RIS is its high array gain, which is made possible by the little noise that passive RISs introduce. Moreover, RIS-empowered systems can be anticipated to significantly increase wireless system capacity using sensing-based awareness to enable transparent 3GPP 5G service \cite{10218358}. Nevertheless, in real-world applications, the anticipated increases in capacity are usually only noticed in wireless communication environments where the primary communication link between the transmitting end and the receiving end is either totally disrupted or extremely weak. Conventional RISs, on the other hand, can only yield minor capacity improvements in many instances where the direct link is not severely attenuated. The path losses of the transmitting end-RIS end and RIS end-receiving end links are multiplied to generate the analogous path loss of the transmitting end-RIS-receiving end wireless link, which is ultimately several times larger than the overall path loss of the directly connected wireless link \cite{zhang2022active}. Consequently, in many wireless situations, passive RISs are nearly incapable to provide significant capacity enhancements. 
While active RISs can dynamically reflect signals with amplified response, they do so at the expense of additional amplifier circuitry cost and energy consumption. This is in contrast to standard passive RISs, which only essentially reflect signals without amplification. Indeed, all RIS-assisted procedures, whether active or passive, necessitate significant logistical assistance and power usage. By evaluating the performance of passive RIS, active RIS, and no RIS-assisted system across a range of transmit power levels, this study aims to determine whether or not RIS-assisted systems are feasible in a challenging radio link condition. \par
In this paper, we propose a three-stage hybrid RIS-assisted algorithm, that can be operated at minimal power consumption in different radio link quality. Moreover, we introduce the concept of inactive RIS alongside the traditional active and passive RIS-assisted systems in a single RIS structure. Our proposed configuration is further verified in part using practical implementation and fabrication of a passive 4.7 GHz RIS-assisted Open Radio Access Network (O-RAN). The rest of the paper is organized as follows. Section 2 describes the related works associated with this research. Section 3 provides the generalized concept of the system model of RIS. Section 4 discusses the fabrication and simulation agenda of the RIS-empowered structure. The results and discussion have been presented in Section 5. Finally, Section 6 concludes this paper.

\section{Related Works}
In the past few decades, wireless communication technology have advanced remarkably. This evolution has taken place across several generations, from 1G to 4G, and is currently undergoing more advancement strategies for 5G and 6G \cite{ratul2023cellular}. RIS have emerged as a promising avenue for wireless communication due to their ability to manipulate wireless propagation. These surfaces consist of sub-wavelength meta-materials and enable precise control over the wireless channel. In the realm of millimeter-wave (mmWave) communication, RIS-assisted systems have the scope to hold the potential for achieving remarkable spectral efficiency while maintaining low hardware complexity and power usage. Consequently, both academic and industrial circles have shown keen interest in RISs for their potential to revolutionize wireless communication. One of the key advantages of RIS technology is its capacity to address coverage challenges in challenging bands like terahertz (THz) and mmWave. By intelligently redirecting wireless signals towards the receiver, RISs can overcome coverage limitations. However, it's imperative to comprehensively assess RIS performance in realistic communication scenarios to ensure their practical viability. In order to validate the performance and feasibility of RIS-assisted systems, several studies have been performed. The related work associated with this research can be broadly divided into three primary categories: passive RIS, active RIS, and hybrid RIS.  
\smallskip
\smallskip
\\
\textbf{Passive RIS-assisted Scenario}\par
Emenonye et al. explored RIS-aided localization, taking into account the impact of RIS misorientation on the positioning error bound (PEB) \cite{emenonye2023ris}. The authors also investigated the effect of RIS orientation offset on RIS orientation estimation. Schroeder et al. conducted an analysis between spectral efficiency (SE) and channel estimation (CE) in mmWave \cite{schroeder2021passive}. A two-stage CE strategy has been proposed for passive RIS and a hybrid RIS design with active components as part of a wave MIMO system. Tradeoffs between hybrid RIS configurations were also investigated. Wang et al. investigated the joint optimization of waveforms design and passive beamforming in RIS-assisted Dual-Functional Radar Communication systems with the goal of minimizing multiple user interference \cite{wang2021joint}. The benefits of combining RIS with passive beamforming are highlighted.
\smallskip
\smallskip
\\
\textbf{Active RIS-assisted Scenario}\par
Lyu et al. demonstrated an active RIS for secure communication within a symbiotic radio system \cite{lyu2023robust}. The authors were able to achieve simultaneous secure delivery to primary and secondary consumers. They introduced a strong secure transmission method, an approach to use active RIS for multicasting, a way to reduce the amount of power used, a strategy to adjust limitations, and an alternating optimization procedure. Active RIS was proposed by Zhang et al. to overcome the capacity constraints of passive RIS under strong direct link conditions \cite{zhang2022active}. They also developed a model for active RIS signals using real-world measurements to validate the claims.
Active and passive RIS were compared by Zhi et al. using the same power budget \cite{zhi2022active}. The ideal power distribution proportions for both types were investigated. It was discovered that active RIS could outperform passive RIS in terms of the signal strength received.
Niu et al. used iterative penalty dual decomposition to build an algorithm for combined optimization of BS precoding and RIS beamforming \cite{niu2023active}. Channel estimation, signal recovery, and optimization of fairness and sum rate were also addressed in their research. 
\smallskip
\smallskip
\\
\textbf{Hybrid RIS-assisted Scenario}\par
Yildirim et al.'s hybrid transmission proposal combines passive RIS with decode-and-forward relay \cite{yildirim2021hybrid}. The authors aimed at maximizing these technologies' interaction for improved performance to maximize efficiency.
A hybrid RIS technique with active and passive elements was presented by Sankar et al. for integrating sensing and communications networks \cite{sankar2022beamforming}. They also discussed the combined design of the RIS coefficients and transmit beamformers.  
Although RIS-assisted systems have been proposed by multiple studies, Trichopoulos et al. developed an experimental RIS prototype and investigated its effectiveness in practical wireless communication environments \cite{trichopoulos2022design}. The authors also assessed RIS performance in outdoor communication environments, focusing on signal-to-noise ratio (SNR) gains using directional antennas. In numerous facets of wireless communication systems, each of these studies expands the understanding of active, passive, and hybrid RIS technologies \cite{zhang2022active}\cite{trichopoulos2022design}\cite{b3}. \par
While several active and passive RIS hybrid structures have been proposed in previous studies, this paper makes an attempt to propose an adaptive, three-stage hybrid RIS-assisted technique that can be operated as an active, passive, or even dormant-state RIS structure. With the goal of lowering power consumption and optimizing operating efficiency in accordance with various radio link quality levels, the inactive state would function exactly like a typical reflector.  This paper also presents the partial results of our fabricated passive RIS-empowered system to validate the feasibility of our approach.

\section{Concept of RIS System Model}

The concept of RIS has been extensively explored in recent studies \cite{schroeder2021passive}\cite{wang2021joint}. In essence, passive RIS units, which have been the focus of most investigations, comprise reflective patches combined with impedance-tunable circuits to facilitate phase modulation as shown in Equation 1. Their operation in a passive manner notably minimizes the impact of thermal noise. However, the anticipated substantial capacity enhancement attributed to passive RISs is frequently unattainable in practical scenarios, particularly when direct links between transmitters and receivers exhibit strong signal strength. This effect stems from the fact that the equivalent path loss of the transmitter-RIS-receiver reflected path is determined by the multiplication, rather than summation, of the individual path losses of the transmitter-RIS and RIS-receiver links \cite{zhang2022activeCONF}. 

\begin{equation}
y = (\mathbf{h}^H + \mathbf{f}^H\underbrace{\boldsymbol{\Phi}^H}_{\text{Phase shift only}}(\mathbf{f}^H\mathbf{G}))\mathbf{w}s + \mathbf{z}
\end{equation}

Equation 1 describes the received signal ($\mathbf{y}$) as a combination of several components. The conjugate transpose of the channel vector ($\mathbf{h}^H$) captures channel effects from BS to UE. The conjugate transpose of the diagonal matrix ($\boldsymbol{\Phi}^H$) incorporates phase coefficients, modeling shifts introduced by reconfigurable surfaces. The operation $(\mathbf{f}^H)$ generates a conjugate transpose of the channel vector between the RIS and UE. Matrix $\mathbf{G}$ signifies the channel matrix between BS and RIS. The beamforming vector ($\mathbf{w}$) shapes the transmitted signal, denoted as $s$, which may carry information. The noise vector ($\mathbf{z}$) accounts for randomness or interference in the received signal. Consequently, this leads to a path loss that is orders of magnitude greater than that of an unobstructed direct link. As a result, in order to accomplish noticeable capacity gain, a large number of RIS elements, frequently in the thousands, are necessary to mitigate for this enormous path loss. \par
To surmount the intrinsic limitations posed by passive RIS, an alternative solution could be found in the form of active RIS \cite{zhang2022activeCONF}. Similar to their passive counterparts, active RIS units possess the capability to manipulate incident signals by altering their phase. Nevertheless, the distinguishing factor is that active RISs can not only modify the phase but also amplify the reflected signals. This is facilitated through the integration of active components, which in turn incurs additional power consumption for signal amplification. It's noteworthy that, unlike passive RIS, the influence of thermal noise introduced by active RIS components cannot be disregarded. Thus, investigating active RIS offers a possible way around the performance constraints inherent in passive RIS. Equation 2 represents the received signal model of active RIS where $\mathbf{P}$ is the amplification matrix
\cite{zhang2022active}.

\begin{equation}
y = (\mathbf{h}^H + \underbrace{\mathbf{f}^H\mathbf{P}\boldsymbol{\Phi}^H\mathbf{G}}_{\text{Amplification}})\mathbf{w}s + \underbrace{\mathbf{f}^H\mathbf{P}\mathbf{n}}_{\text{Extra noise from active RIS}} + \mathbf{z}
\end{equation}

Although passive RISs have an inherent benefit when it comes to reducing thermal noise, active RISs are distinguished by their capacity to amplify reflected signals, even though this comes with a higher power consumption and the addition of thermal noise to the system model. Fig. 1 depicts the basic functionality of active and passive RIS.

\begin{figure}
    \centering
\includegraphics[width=\linewidth]{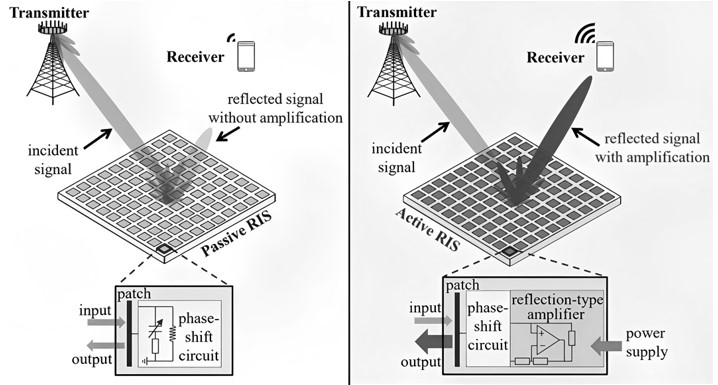}
    \caption{Hardware setup and beamforming of a passive and an active RIS. \cite{zhang2022active}.}
\end{figure}

\begin{algorithm}
\caption{Proposed Three Stage Hybrid RIS System}
\label{algo:prop_algorithm}
\begin{algorithmic}[1]
   \STATE \textbf{Initialization:} \( \mu \) symbolizes the \textbf{\textit{measurement\_report}} messages: \( \mu_{\text{W}} \) (weak transmit power), \( \mu_{\text{S}} \) (strong transmit power), and \( \mu_{\text{H}} \) (high transmit power \( > 60 \, \text{dBm} \)).
   \STATE \textbf{Input:} Channel conditions \( h_k \) for \( k \in \bigcap_{k=1}^n \{k\} \), threshold transmit power \( \rho \), gain \( \tau \).
   
   \FOR{each \( k \)}
        \STATE Determine \( \mu_k \) based on \( h_k \).
        
        \IF{\( \mu_k = \mu_{\text{W}} \) \textbf{(\textit{measurement\_report\_weak})}}
            \STATE \textbf{Enhance Signal Strength:} (Active RIS if \( \tau > \rho \), else Select Passive RIS)
            
        \ELSIF{\( \mu_k = \mu_{\text{S}} \) \textbf{(\textit{measurement\_report\_strong})}}
            \STATE \textbf{Enhance Signal Strength:} (Active RIS if \( \tau > \text{passive RIS gain} \), else Select Passive RIS)
            
        \ELSIF{\( \mu_k = \mu_{\text{H}} \) \textbf{(\textit{measurement\_report\_high})}}
            \STATE \textbf{Avoid Active RIS:} (Switch to Passive RIS if \( \tau > \rho \), else Power Off and Initiate Dormant RIS)
            
        \ENDIF
   \ENDFOR
\end{algorithmic}
\end{algorithm}

\begin{figure*}

    \centering
    \includegraphics[width=0.98\linewidth]{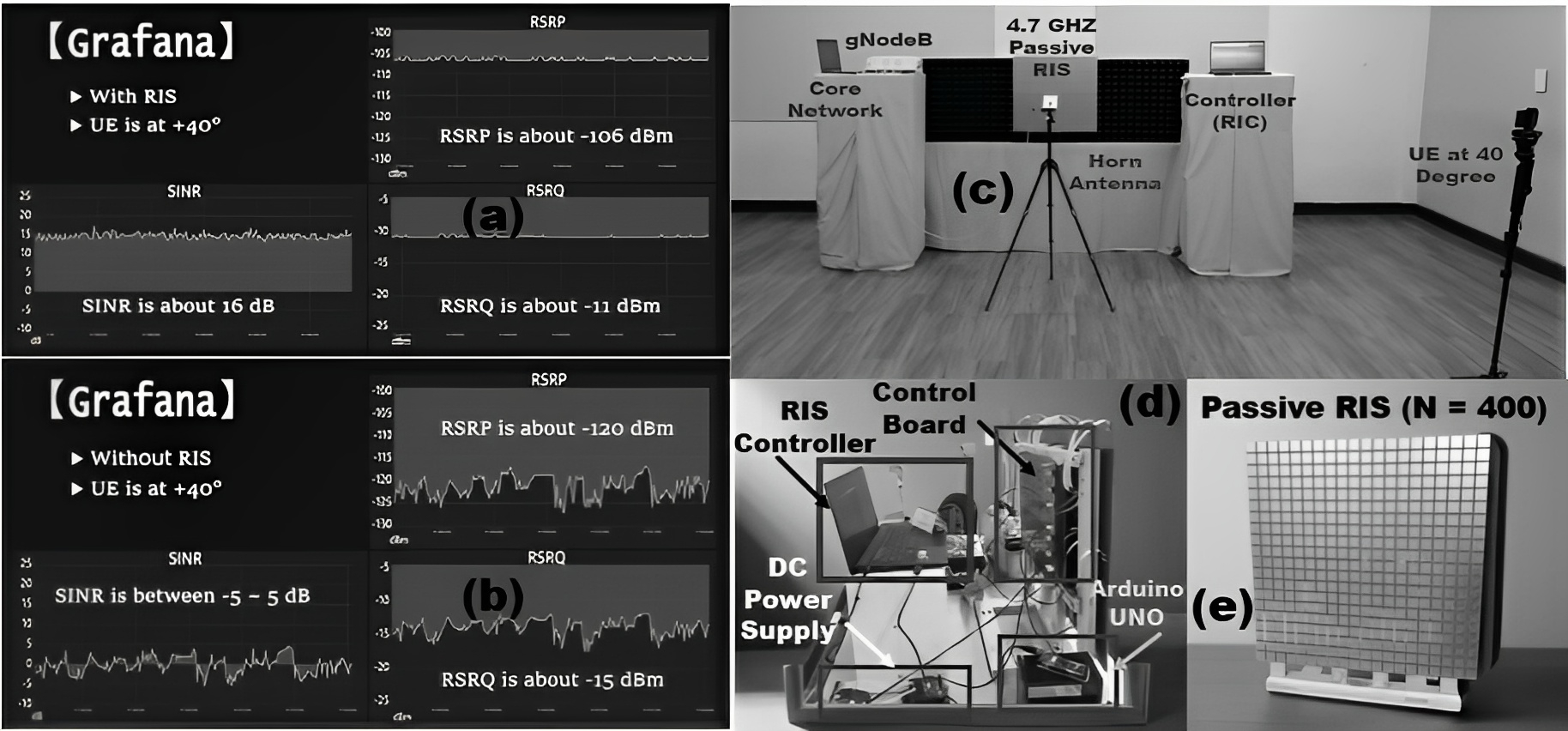}
    \caption{Experimental passive RIS (a) with RIS (b) without RIS (c) overall RIS-assisted environment (d) behind the radiation layer (e) fabricated RIS}
    
\end{figure*}

\section{Fabrication and Simulation Setup}

To validate the proposed three-stage hybrid RIS configuration, firstly, we have designed and fabricated a passive RIS structure with  $ \pm 40^\circ $  phase shifting capacity for experimental measurements, as shown in Fig. 2. Note that this design is in the primary stage of development with 400 arrays and can be applied directly to the case of large arrays. Furthermore, the phase-shifting capability of passive RIS has been thoroughly demonstrated as shown in Fig. 2 (a) and Fig. 2 (b). Integrating our proposed simplified algorithm as shown in Algorithm 1 for adaptive hybrid configuration with the necessary module has the potential to perform equally well. Fig. 2(c), Fig. 2(d), and Fig. 2(e) depict the experimental setup of the RIS-assisted environment, the opposite portion of the RIS radiation layer, and the main fabricated RIS structure. However, as the active RIS fabrication was not available during this study, we explore a system augmented by an active RIS in \cite{zhang2022activeCONF}. It needs to be noted that, for keeping the simulation environment simplified and fair, the fabricated passive RIS precoding has not been implemented in the simulation. Rather, for the active RIS, we utilize the algorithm proposed in \cite{zhang2022activeCONF} for the joint design of precoding and beamforming. Meanwhile, for passive RIS, we adopt the methodology outlined in \cite{pan2020multicell}.

In particular, we investigate two distinct scenarios marked by differing channel conditions. In the first scenario, the direct link stands strong due to favorable radio link quality or unimpeded signal propagation between the UE and the BS. Conversely, in the second scenario, the direct link exhibits weakness. To be precise, we employ two distinct path loss models from the 3GPP TR 38.901 standard to describe the significant attenuation of the channels\cite{3gpptr38901}:
\begin{align}
    PL_{\text{strong}} &= 13.54 + 39.08 \log d \\
    PL_{\text{weak}} &= 37.3 + 22.0 \log d
\end{align}

Here, $d$ signifies the distance between the two devices. $PL_{\text{weak}}$ characterizes the feeble BS-UE link in scenario 2, whereas $PL_{\text{strong}}$ models the robust BS-UE link in scenario 1. It's notable that, for both scenarios, $PL_{\text{weak}}$ is employed to replicate the characteristics of the BS-RIS and RIS-UE channels, creating an intentionally challenging overall radio link condition. Incorporating the effects of small-scale fading, we adopt the Rician fading channel model for all relevant communication channels.

The BS is positioned at coordinates $(0, -45 \, \text{m})$, and the three-stage hybrid RIS-assisted structure is located at $(180\, \text{m}, 20 \, \text{m})$. Additionally, five users are randomly distributed within a circular region with a radius of 6 m, centered at coordinates $(200 \, \text{m}, 0)$. The configuration involves 5 BS antennas and 400 RIS elements ($N = 400$). The ambient noise power is set to -65 dBm. In order to facilitate an equitable comparison, we impose a constraint on the total power consumption. As a result, the collective transmit power for the active RIS encompasses the power utilized by both the active RIS itself and the BS. Importantly, considering that the power consumption of passive RIS is significantly lower than that of active RIS, only the power associated with the BS is taken into account when calculating the passive RIS's total transmit power. Similarly, for dormant RIS, for the sake of simplicity, we assume power consumption similar to that of passive RIS, although in practical scenarios, the energy consumption would indeed be lower.

\section{Result and Discussion}

Though RIS-assisted beamforming can significantly improve user experience in terms of signal gain, its effectiveness is not consistent in all situations. Excessive directional transmit power has a negative impact on the performance of the active RIS-assisted system. Moreover, passive RIS or even no RIS performs better at transmit powers greater than 60 dBm.
\begin{figure} 
    \centering
    \includegraphics[width=\linewidth]{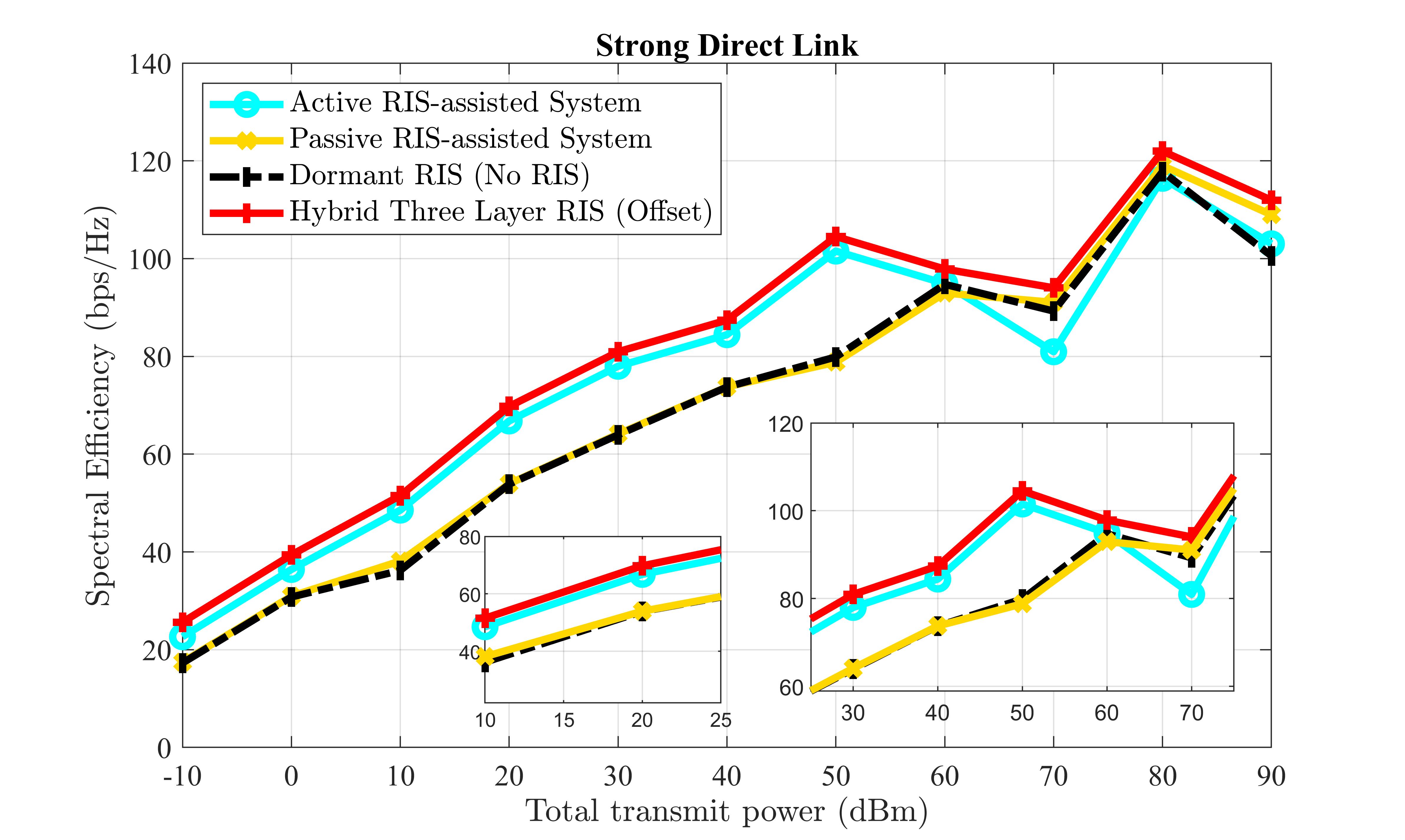}
    \caption{Performance comparison of RIS with strong direct link.}
    
\end{figure}

\begin{figure}
    \centering
    \includegraphics[width=\linewidth]{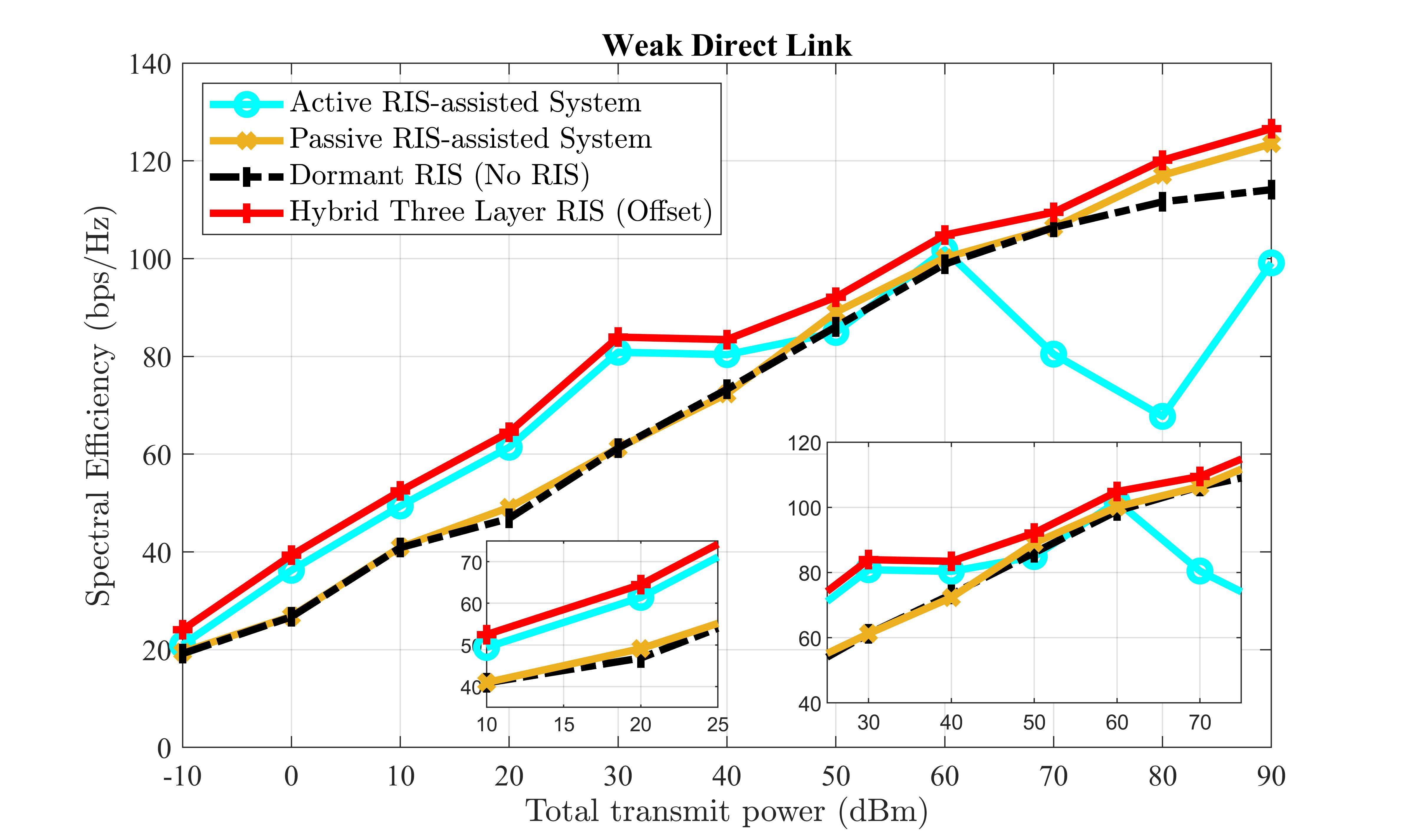}
    \caption{Performance comparison of RIS with weak direct link.}
    
\end{figure}
Fig. 3 and Fig. 4 illustrate the relationship between spectral efficiency and total power consumption for the two distinct scenarios under examination, characterized by strong and weak direct links respectively. In the initial scenario, where a robust direct link is present (scenario 1), the passive RIS exhibits marginal performance gains, whereas the active RIS demonstrates comparatively superior results. It is important to acknowledge that the BS-RIS and RIS-user channels are designed with a notably weak path loss model. This is why, despite the inclusion of the passive RIS, the overall performance remains relatively modest. That is why, in cases where path loss models and the broader radio link quality are conducive to efficient wave propagation, our devised passive RIS outperforms the passive RIS presented in the simulation. This difference in performance stems from the use of distinct algorithms, channel models, and propagation environments between the simulated and fabricated passive RIS setups.
The simulation results reveal that while active RIS-assisted systems perform better in both scenarios, increasing transmit power beyond 60 dBm leads to a significant decrease in spectral efficiency, making active RIS less viable for maintaining user experience. During high transmission power and significant power output, the passive RIS outperforms the active RIS. High power transmission levels are uncommon in conventional deployments, but might be relevant in future intelligent deployments.

\section{Conclusion}
This paper presents a simple explanation of the principles of passive and active RIS. It carries out an extensive comparison study that includes dormant, passive, and active RIS. Additionally, the work proposes a compact three-stage hybrid RIS method for best spectral efficiency. A practical implementation of 4.7 GHz passive RIS-assisted O-RAN, with a $20 \times 20$ array, has been included in the scope of this investigation. Under real-world conditions, this implementation validates the phase modulation in the initial phase of our proposed strategy. Under real-world conditions, this implementation validates the phase modulation in the initial phase of our proposed strategy. The simulation phase, on the other hand, employs separate precoding methods for RIS testing, replicating a discerning outdoor scenario with higher path loss. The software simulation clarifies the effects of poor radio link quality. Results from the study confirm that active RIS outperforms passive RIS in both standard and dynamic wave propagation scenarios. However, active RIS efficacy degrades significantly when total transmission strength exceeds 60 dBm. This decline might be linked to the increased additional noise that occurs with the increased transmission power of active RIS.

\section*{Acknowledgment}
      
This work was supported in part by the Advanced Institute of Manufacturing with Hightech Innovations (AIM-HI) from the Featured Areas Research Center Program within the framework of  the Higher Education Sprout Project by the Ministry of Education (MoE) in Taiwan.

\end{document}